\documentclass{ab2018}
\usepackage{url}
\usepackage{graphicx}
\usepackage{natbib}
\usepackage{graphicx}
\usepackage{eqnarray}
\usepackage{amsmath}
\usepackage{amssymb}

\begin{document}

\title{Measurements of the Expansion Velocities of Ionized-Gas Superbubbles
in Nearby Galaxies
Based on Integral Field Spectroscopy Data}

\author{Grigorii~V. Smirnov-Pinchukov$^1$,  Oleg~V.~Egorov$^{2,3}$}

\institute{
$^1$ Max Planck Institute for Astronomy,
D-69117 Heidelberg, Germany \\
$^2$ Astronomisches Rechen-Institut,
Zentrum f\"ur Astronomie der Universit{\"a}t Heidelberg, D-69120 Heidelberg, Germany \\
$^3$ Sternberg Astronomical Institute, M.V. Lomonosov Moscow State University, Moscow, 119234 Russia
}

\titlerunning{Expansion velocities of superbubbles in IFU data}

\authorrunning{Smirnov-Pinchukov \& Egorov}

\date{April 30, 2021/Revised: September 6, 2021/Accepted: September 6, 2021}
\offprints{\email{smirnov@mpia.de} (GVS)}

\abstract{
The study of dynamic properties of bubbles in the
interstellar medium is important for understanding the feedback
mechanisms from star-formation processes in galaxies. The ongoing integral field
spectroscopy of nearby star-forming galaxies reveals many expanding
bubbles and superbubbles identified by the local increase in gas velocity
dispersion. The limited angular resolution often prevents bona fide
measures of the expansion velocities in galaxies outside the Local
Group, even despite sufficient spectroscopic resolution. We present a
method that makes it possible to measure the expansion velocity of
bubbles surrounding massive stars and clusters based on the data about
local variations in gas velocity dispersion. We adapt the method for the
Fabry--Perot interferometers used with the 6-m telescope of the Special
Astrophysical Observatory of the Russian Academy of Sciences, as well as
for any  integral field spectrograph with a Gaussian line spread function. We
apply the method described to analyze the kinematics of ionized superbubbles
gas and the only known supernova remnant in the IC\,1613 galaxy. The
estimate of the kinematic age of the supernova remnant (on the order of
3100~years) agrees well with the previously obtained independent
estimate based on X-ray data.
}

\maketitle

\section{INTRODUCTION} \label{intro}

Stars with masses $M>8~M_\odot$ play an important role in the evolution
of the interstellar medium in galaxies. Their wind, ionizing radiation,
and subsequent supernova explosions are the source of energy for the
formation of shell-like structures surrounding clusters and OB-associations.
The sizes of the observed bubbles vary over a wide range---from several
parsecs to several hundred parsecs in the case of supershells/superbubbles
\citep{Oey1997, Nath2020}. Sustained energy inflow from several
generations of stars results in further growth of superbubbles to sizes
of 1--3~kpc \citep{Weisz2009a, Warren2011}. In some dwarf galaxies such
giant superbubbles are the dominant structure of the interstellar medium
\citep{Egorov2018}. Such superbubbles that are larger than a few hundred
parsecs are mostly observed in H\,I at 21~cm \citep{Bagetakos2011},
although deep images in emission lines allow detection of similar
structures in ionized gas \citep{Egorov2014, Egorov2017}.

The classical model of the evolution of bubbles driven by the
winds of massive stars \citep{Weaver1977} and the combined wind of multiple
stars and supernova explosions \citep{MacLow1988} describes the observed
structures in the interstellar medium of galaxies quite well at
qualitative level. According to this model, the radius of a bubble $R$ depends
on time $t$ and the inflow of mechanical energy $L$ as $R \sim
L^{1/5}t^{3/5}$. Accordingly, at each moment of time its  expansion
velocity is $v_{\rm exp} \sim L^{1/5}t^{-2/5}$. Thus the age of the
bubble and the energy required for its formation can be estimated based
on the observed quantities---the size and expansion velocity. In
particular, the age of the shell can be estimated in terms of the
\citet{Weaver1977} model as \mbox{$t=0.6R/v_{\rm exp}$}. Because of
this rather simple analytical dependence the observed parameters of
superbubbles in nearby galaxies can serve as indicators of the properties
of massive stars responsible for the formation of these structures.
However, a numerical comparison of observed superbubbles properties with
their estimates predicted by the \citep{Weaver1977} model showed a
severalfold-factor discrepancy \citep{Oey2004}. Radiative energy losses,
which can range from 60\% to almost 100\% \citep{Sharma2014,
Vasiliev2015a, Yadav2017}, are viewed as one of the main causes of this
discrepancy. The new analytical model of superbubble evolution proposed
by \citet{El-Badry2019} takes this factor into account as a parameter.
However, it remains an open question how the efficiency of energy
transfer from star clusters into the mechanical energy of expanding superbubbles
changes under different conditions (e.g., in the case of different
metallicities and gas densities).

The development of integral field spectroscopy techniques has made possible a
detailed comparison of observed properties of massive stars and of the
surrounding regions of ionized gas in several nearby galaxies, allowing
the contribution of various feedback processes to the overall process of
energy and momentum exchange with the interstellar medium to be
estimated \citep{Egorov2014, Egorov2017, McLeod2019, Ramachandran2019,
McLeod2020}. However, such studies are significantly limited by the
spatial and spectroscopic resolution of observational data. Thus with
the characteristic sizes of the shells surrounding individual massive
stars or small OB-associations on the order of several tens of parsecs,
even at a distance of 4--5~Mpc (e.g., the M\,81 group \citep{Karach2013}) 
the angular sizes of such objects often become comparable to the
angular resolution of observational data (on the order of
$2\arcsec$--$3\arcsec$), making the determination of superbubbles sizes and
expansion velocities quite a challenging task. Furthermore, the
characteristic observed superbubbles expansion velocities
\mbox{20-50}~km\,s$^{-1}$ make it impossible for most modern integral field
spectrographs to resolve the emission line profiles into components to
reliably estimate their expansion velocities. Observations of a number
of nearby dwarf galaxies made with the Fabry--Perot interferometer (FPI)
in the H$\alpha$ line demonstrate the presence of a substantial number
of regions of increased ionized gas velocity dispersion whose position
on the ``intensity--velocity dispersion'' diagram ($I$--$\sigma$)
allows us to interpret them as spectroscopically unresolved expanding
superbubbles \citep{Moiseev2012}. Numerical simulations of the
interaction of multiple supernova remnants yield a close to the observed
pattern in the ($I$--$\sigma$) parameter space, which confirms the above
interpretation \citep{Vasiliev2015}. As shown by \citet{Guerrero1998},
such superbubbles should exhibit radial gradient of the emission line
profile width, which can be used to estimate the expansion velocity,
however, such measurements require a rather high angular resolution. On
the other hand, \citet{TT1996} showed that the presence of a spatially
unresolved shell also broadenes the observed emission line profile. In
particular, \citet{Moiseev2012} proposed to use this effect to search
for unique emission objects with high stellar wind speeds (e.g.,
Wolf-Rayet stars, luminous blue variables) on the ($I$--$\sigma$)
dependence for the galaxies studied. Thus, even in cases of insufficient
spatial and spectroscopic resolution, expanding bubbles around stars and
clusters can be identified by changes in the width of the spectral line
profile. However, the physical parameters of such  bubble (in
particular, its expansion velocity) often cannot be estimated directly
without invoking model assumptions about its geometry.

In this paper, we propose a method that allows us to estimate the
expansion velocity of the bubbles and superbubbles in the interstellar medium from
measurements of the emission line profile width in their integrated
spectra and from the average velocity dispersion of the interstellar
medium in the galaxy. The method is based on the analytical model of a
uniform spherically symmetric expanding shell in the turbulent
interstellar medium, however we show that the estimates are also valid
for the case of non-uniform shells, and agree well with ``direct''
measurements of the expansion velocities and ages of observed superbubbles
and supernova remnants (based on the example of the IC\,1613 galaxy). The
method described above is applicable, in particular, to spatially
unresolved (or poorly resolved) shells, as well as in the cases where
spectroscopic resolution prevents decomposition of the line profile into
kinematically isolated components. In this paper we consider two types
of spectrographs: with an instrumental profile (line spread function,
LSF) in the form of a Lorentz function (as in FPI) and with a Gaussian instrumental profile
(applicable to the case of many classical integral field spectrographs). We have
considered Lorentzian LSF widths corresponding to the FPIs
used\footnote{Actual list:
\url{https://www.sao.ru/hq/lsfvo/devices/scorpio-2/ifp_eng.html}} in
SCORPIO-2 \citep{scorpio2} focal reducer of the 6-m telescope of the
Special Astrophysical Observatory of the Russian Academy of Sciences (SAO RAS)
\citep[see review of the method in][]{Moiseev2021}.

This paper has the following layout. Section~\ref{sec:model_bubble}
describes the analytical model of a uniform expanding bubble in a
turbulent medium and the factors that affect the shape of its integrated
emission line profile (turbulence of the interstellar medium, thermal
and instrumental broadening), and provides a qualitative comparison of
the constructed model with the H$\alpha$ line profiles of observed
objects. Section~\ref{sec:method_apply} describes the proposed method
for measuring the expansion velocity of a bubble from the observed
velocity dispersion, and the results of its Monte Carlo testing. In
Section~\ref{sec:obs}, we use the method described above to estimate the
expansion velocity of the expanding superbubble of ionized gas in
IC\,1613 and determine the kinematic age of the only supernova remnant
in this galaxy. We summarize the main conclusions of this work in
Section~\ref{sec:summary}.

\section{MODELING THE PROFILE OF AN EXPANDING BUBBLE}\label{sec:model_bubble}
\subsection{A Uniform Expanding Bubble in Unperturbed Interstellar Medium}\label{sec:model}

\begin{figure}
\includegraphics[scale=0.4]{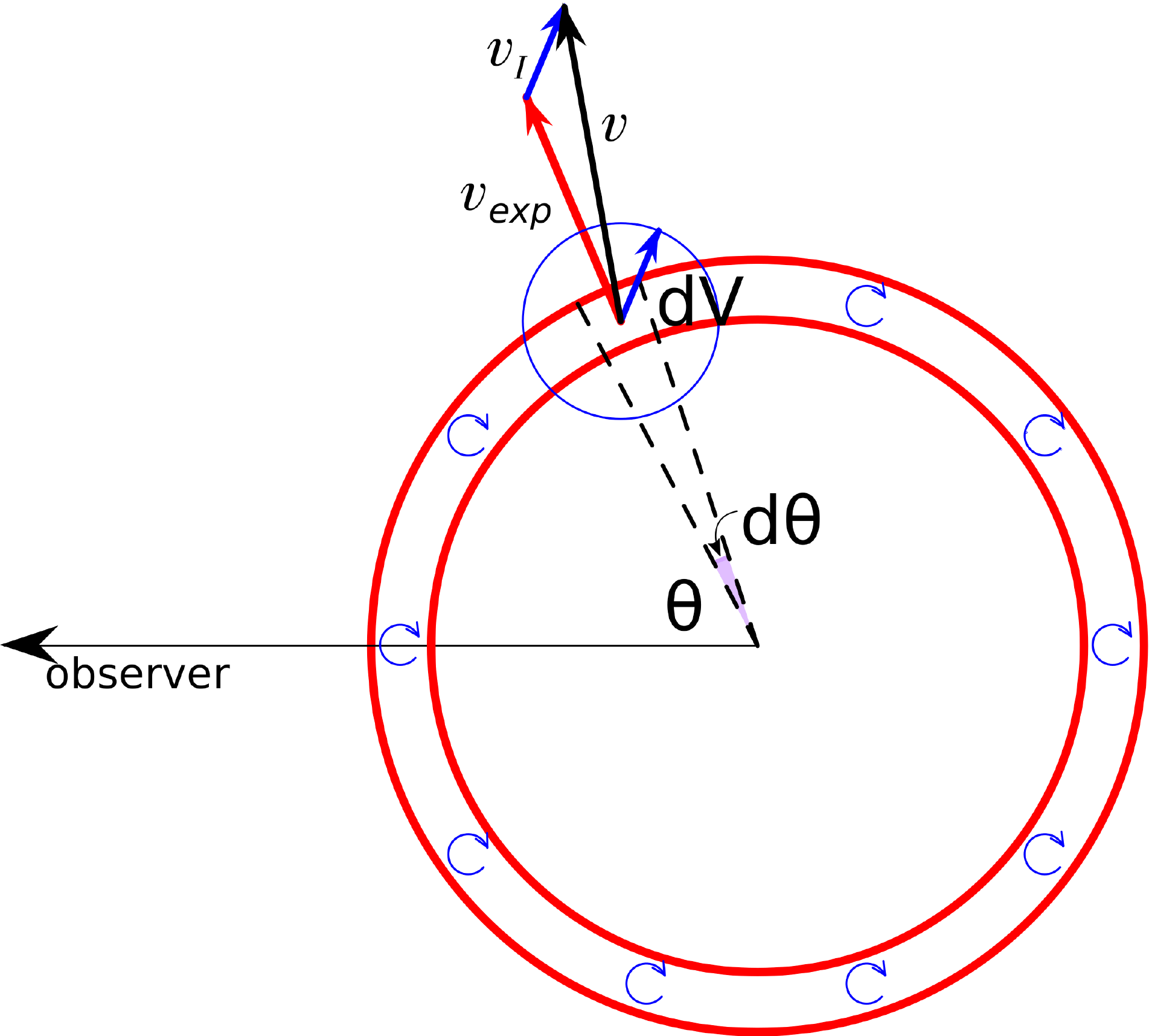}
\caption{Scheme of a uniform expanding bubble model. The velocity of an
emitting atom from spherical layer $dV$ with polar angle $\theta$
relative to the center of the shell is equal to the sum of the bubble
expansion velocity $v_{\rm exp}$ (red) and the random velocity of
thermal and turbulent motions $v_I$ (blue) that causes the emission line
broadening.}  \label{fig:isosphere} \end{figure}

According to the classical model of the evolution of a bubble whose
expansion is driven by the energy inflow from the wind of massive stars
and supernovae \citep{Weaver1977, Lozinskaya1992}),
propagation of shocks in the interstellar medium results in the
formation of a rather thin layer of swept-up matter with the same gas
pressure inside this layer. According to formula~(73) from
\citet{Weaver1977}, in the case of characteristic bubble expansion
velocities \mbox{$v_{\rm exp}\sim20$--$50$~km\,s$^{-1}$} and the sound
speed in ionized gas equal to $c_s\sim10$~km\,s$^{-1}$ the thickness of
the newly formed bubble equals to 2--7\% of its radius. The actually
observed shells are usually somewhat thicker (e.g., 7--25\% of the
radius in the LMC \citep{Oey1996}, on average about 25\% for infra-red bubbles
in the Galaxy \citep{Churchwell2006}) but their thickness is still small
compared to the diameter, which allows us to use the uniform thin
sphere model to describe observed gas kinematics toward it (see
Fig.~\ref{fig:isosphere}).

Consider a segment of the spherical layer $(\theta, \theta+d\theta)$ of
gas with volume $dV$, whose radial velocity is concentrated in the
interval $(v, v+dv)$, where $v$ is determined exclusively by the
expansion velocity of the bubble as \mbox{$v=v_{\rm sys}+v_{\rm
exp}\ \theta$}. Here $v_{\rm sys}$~is the radial velocity of the
bubble center and $\theta$~is the angle between the direction to the
observer and the normal vector to the surface element (polar angle). The
radial velocity differential for such an element is \mbox{$dv=-v_{\rm
exp}\sin{\theta}d\theta$}. We assume that the thickness of the shell is
constant and equal to $h$ to obtain

\begin{equation}
\begin{array}{rcl}
 dV&=&\int\limits_0^{2\pi} \left(h r^2\sin{\theta} d\theta\right)
d\varphi\\ &=&2\pi hr^2 \sin{\theta} d\theta\\\\ &=& -2\pi \dfrac {hr^2}
{v_{exp}} dv,
\end{array}
\end{equation}
where $r$~is  the radius of the shell and $\varphi$ is the azimuthal
angle in the plane perpendicular to the sky plane (see
Fig.~\ref{fig:isosphere}). It follows from this that

\begin{equation}
\frac{dV}{dv}=-2\pi \frac {hr^2}{v_{\rm exp}}=const.
\end{equation}
The radiation flux toward the observer from each element $dV$ is
proportional to the number of radiating atoms in it
\citep{Osterbrock2006}. Assuming the density of matter is the same
throughout the shell, we obtain
\begin{equation}
F_v(v)dv \propto dV = const\times dv.
\end{equation}

\begin{figure*}
\centering
\includegraphics[scale=0.5]{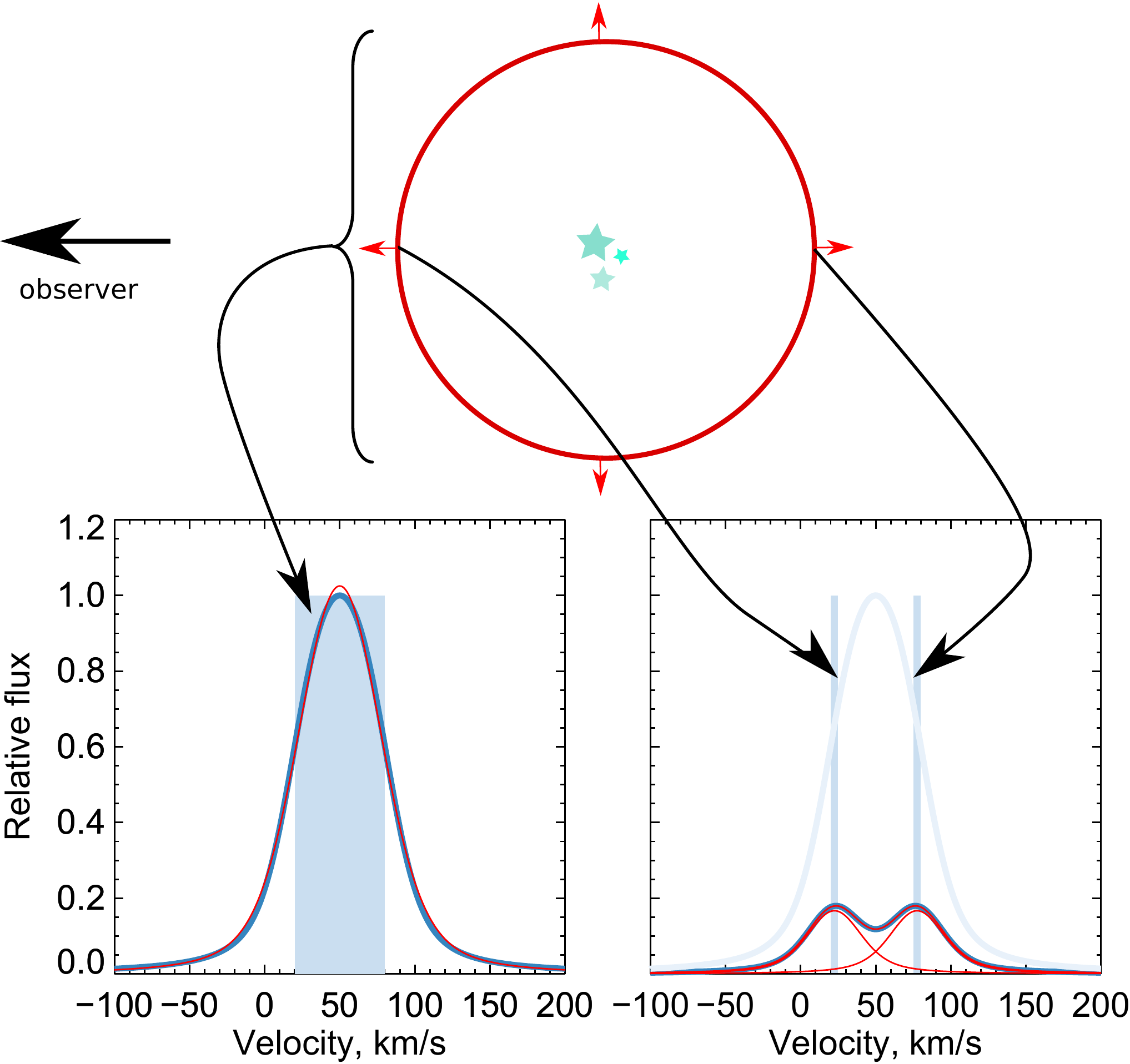}
\caption{Schematic representation of the expanding shell (top) and
examples of the emission-line profiles in its different parts (bottom).
Bottom left: integrated emission-line profile of the shell, bottom
right: line profile towards the center of the shell. The blue rectangle
shows the model spectrum excluding thermal, turbulent, and instrumental
broadening. The blue curve shows the line profile including thermal,
turbulent, and instrumental broadening, the red line shows the one- or
two-component Voigt profile fit.} \label{fig:nonturb_profile}
\end{figure*}

Thus in the uniform bubble model considered the
observed integrated line profile has a rectangular shape with the same
radiation flux density $F_v(v)$ over the velocity interval $[v_{\rm
sys}-v_{\rm exp}; v_{\rm sys}+v_{\rm exp}]$ also in the absence of
radiation outside this interval (see the left-hand panel in
Fig.~\ref{fig:nonturb_profile}). In contrast, if only the central part
of such an idealized model is observed, the emission line-profile has
the form of two narrow rectangular peaks with velocities corresponding
to the ray velocities of the approaching and receding sides of a bubble, and
their difference is equal to $2\times v_{\rm exp}$ (see the right-hand
panel in Fig.~\ref{fig:nonturb_profile}).

In the above scheme only the contribution of the bubble
expansion to the line profile is taken into account, ignoring the
thermal and natural broadening and the contribution of the turbulence of
the interstellar medium. Let us consider these effects in the next
section.

\subsection{Estimating Thermal and Turbulent Line Broadening}
\label{sec:fwhm}

The width of emission lines in the interstellar medium is influenced by
three main factors: natural line broadening (due to quantum uncertainty of
the level energy with a finite lifetime), thermal line broadening and
gas turbulence. The contribution of natural line broadening in this case
is negligible compared to other factors, so we will consider only
the effect of the latter two factors.

Under the assumption of local thermodynamic equilibrium, the velocity of
thermal motion of electrons in H\,II-regions is described by the Maxwell
distribution, according to which the dispersion of the velocity
projection is equal to
\begin{equation}
\sigma_{\rm{th}}=\sqrt{\dfrac{k_B T_e}{m}},
\end{equation} which corresponds to $\sigma_{\rm th}\sim
9.1$~km\,s$^{-1}$ for HII for the adopted characteristic electron
temperature $T_e=10^4$~K \citep{Osterbrock2006}

The situation with estimating the velocity dispersion of turbulent
motions in the interstellar medium is more complicated. Observations
indicate supersonic velocity dispersion for turbulent motions of ionized
gas in galaxies \citep{Moiseev2015turb, Varidel2020}, which may be due
to gravitational instability combined with feedback from massive stars
\citep{Krumholz2018}. The observed dispersion of ionized gas velocities
in this case correlates with the star-formation rate in the galaxy, and
the characteristic velocity dispersions in nearby dwarf galaxies are
$\sigma_{\rm turb} \sim 12$--$35$~km\,s$^{-1}$ \citep{Moiseev2015turb},
although they can be several times greater for galaxies with high
star-formation rates. With integral field spectroscopy data available this
quantity can be estimated for each individual object under study, for
example, as the flux-weighted average of the velocity dispersion of
bright H\,II regions \citep{Moiseev2012} in the H$\alpha$ line.

Thus because of thermal broadening and turbulent motions in the
interstellar medium the observed emission lines originating from outside
the shell have a Gaussian profile with velocity dispersion
\begin{equation}
\sigma_{\rm ISM}=\sqrt{\sigma_{\rm th}^2+\sigma_{\rm turb}^2}\sim
15\hbox{--}36~\text{km\,s$^{-1}$}
\end{equation}

Assuming that the above mechanisms produce the comparable broadening
of the emission lines from the walls of a bubble and
from the surrounding interstellar medium, the final line
profile from the shell can be obtained by convolving the rectangular
distribution obtained in the previous section with a Gaussian function
with a width of $FWHM=2\sqrt{2\ln{2}}\,\sigma_{\rm ISM}$.

\subsection{Comparison of Model and Observed Line Profiles}
\label{sec:model_lsf}

The observed shape and width of a spectral line depend on the
instrumental profile of the spectrograph. The instrumental profile of
the FPI in SCORPIO-2 is well described by the Lorentz function, which
means that the final observed emission-line profile can be approximated
by the Voigt function, which is a convolution of the Gaussian and
Lorentz functions \citep{Moiseev2008}. Fig.~\ref{fig:nonturb_profile}
shows that after thermal, turbulent, and instrumental broadening are
taken into account, the line profile from the model considered is indeed
similar to those observed with the FPI (see Section~\ref{sec:obs}). The
integrated line profile can be well described by a single-component Voigt function, and
the line profile in the direction of the  center of a bubble, by a
two-component Voigt function with the components centered on the
velocities of the approaching and the receding sides.

\begin{figure}
\includegraphics[scale=0.6]{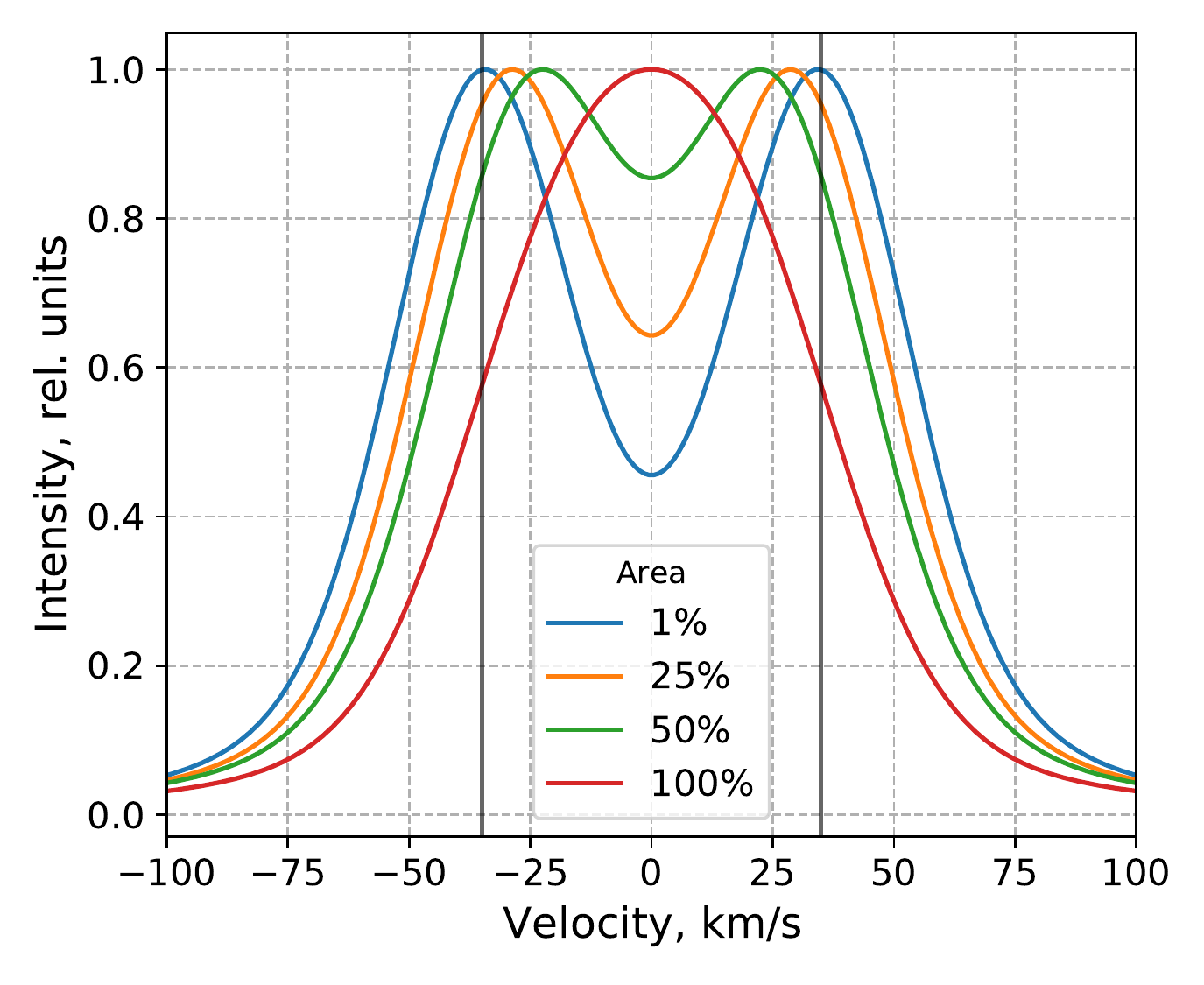}
\caption{Model emission-line profile of a bubble expanding with velocity
$v_{\rm exp}=35$~km\,s$^{-1}$ (marked with vertical lines) in the medium
with \mbox {$\sigma_{\rm ISM}=15$}~km\,s$^{-1}$ and ``observed'' with
spectrograph with Lorentzian LSF \mbox {$FWHM_{\rm
LOR}=18$}~km\,s$^{-1}$. Different colors correspond to different
apertures in the integration: 1\% of the total area in the sky plane
corresponds to the center of the bubble and 100\%, to the entire shell.}
\label{fig:appertures} \end{figure}

We now use the model constructed to illustrate how insufficient angular
resolution can distort the estimate of the bubble expansion velocity
determined by decomposing the observed profile into components.
Fig.~\ref{fig:appertures} shows examples of simulated shell line
profiles (expanding with a velocity of $v_{\rm exp}=35$~km\,s$^{-1}$ in
a medium with $\sigma_{\rm ISM}=15$~km\,s$^{-1}$ and observed with an
IFP751 FPI having LSF width $FWHM_{\rm LOR}=18$~km\,s$^{-1}$) integrated
over different apertures. The smallest aperture corresponds to the
central part of the shell, and the largest, to the integrated spectrum
from the entire region. While in the former case the line profile
clearly breaks into two components, there is no such separation in the
case of the integrated spectrum corresponding to the case of a
completely unresolved shell. The separation between the individual
components can be seen to decrease with increasing relative area of the
aperture compared to the total area occupied by a bubble in the sky plane. In the
case considered, the shell expansion velocity is underestimated by a
factor of 1.5 for an aperture with 50\% of the total bubble area. Thus
even if spectroscopic resolution is sufficient to decompose the profile
into kinematically distinct components, insufficient angular resolution
may result in a significantly underestimated shell expansion velocity
and hence overestimated kinematic age and underestimated energy inflow
required for the formation of  these components. To correctly determine
these parameters the angular resolution of the data has to be
0.5 of the shell size or better. This imposes a significant limitation
on the use of estimates obtained by line-profile decomposition for
bubbles smaller than 100~pc in the case of observations of galaxies
beyond the Local Group.

In addition to the decomposition of the observed profile into
kinematically isolated components, there are a number of other methods
for estimating the expansion velocities of nebulae from published
observational data. \citet{Guerrero1998} suggested using radial gradient
of the line profile width to compute the expansion velocities of
nebulae. However, this method is not applicable in the case of
insufficient angular resolution or nonuniform nebulae. \citet{Dewey2010}
described yet another method in his paper on the interpretation of the
kinematics of unresolved young supernova remnants. Like in this paper
(cf. Section~\ref{sec:model}), the above author assumes that the
integrated spectrum of an expanding supernova remnant has a uniform
distribution with velocities from \mbox {$-v_{\rm exp}$} to $v_{\rm
exp}$, but he ignores the effects due to turbulence in the surrounding
interstellar medium (which is true in first approximation given that
characteristic shock velocities in young supernova remnants are one to
two orders of magnitude  higher than those discussed in our paper) or
the non Gaussian shape of the spectrograph instrument profile. Dewey
computes the dispersion of the uniform distribution from $a$ to $b$
(from \mbox{$-v_{\rm exp}$} to $v_{\rm exp}$) as
\begin{equation}
\sigma = \sqrt{\dfrac{(b - a)^2}{12}} = \sqrt{\dfrac{v_{exp}^2}{3}}
\approx 0.577 \ v_{\rm exp},
\end{equation}
to determine the expansion velocity of supernova remnants from the
observed line width. Note that this is not entirely correct, because the
velocity dispersion estimated by fitting the distribution to a Gaussian
function ($\sigma \approx 0.715\ v_{\rm exp}$) differs from that given
above for the uniform distribution. With the above fact taken into
account, the solution proposed by \citet{Dewey2010} coincides with a
special case of the method we consider below (see
Section~\ref{sec:method_apply}) in the case of zero turbulent motions in
the interstellar medium.

Thus, simple approaches can yield significantly (by tens of percent)
underestimated or overestimated results. Below we propose a method for
estimating shell expansion velocities from the measured width of the
integrated line profile based on modeling that takes into account the
main effects described above that affect the observed shape of the
profile.

\section{MEASURING A BUBBLE EXPANSION VELOCITY BASED ON THE OBSERVED
VELOCITY DISPERSION} \label{sec:method_apply}

As noted above, when the angular and/or spectral resolution of the
employed is insufficient, the expansion of a bubble shows
up as an increase in the measured velocity dispersion. This effect is
often apparent in observations. Thus one can efficiently search for
expanding superbubbles in nearby galaxies by analyzing ($I$--$\sigma$)
diagrams in the H$\alpha$ line \citep{MunozTunon1996, Moiseev2012,
Egorov2017, Egorov2018, Egorov2021}, where such objects can be distinguished by
their increased velocity dispersion. Compact nebulae around individual
energetic objects (WR, LBV stars) and unresolved supernova remnants also
stand out on the ($I$--$\sigma$)-diagram as regions with high intensity
and high velocity dispersion \citep{Moiseev2012}. Below we investigate
how to estimate the expansion velocities of such bubbles based on the
data about the spatial distribution of velocity dispersion in their host
galaxy.

To reveal the relationship pattern between the measured spectral line
width and the actual shell expansion velocity, we constructed and
analyzed a series of model spectra computed for the case of a uniform
bubble expanding in turbulent interstellar medium (see
Section~\ref{sec:model_bubble}). In our analysis we varied the shell
expansion velocity and the velocity dispersion in the interstellar
medium (which includes turbulent and thermal line broadening) in the
intervals \mbox {$v_{\rm exp}=5$--$95$}~km\,s$^{-1}$ and $\sigma_{\rm
ISM}=3$--$70$~km\,s$^{-1}$, respectively. As mentioned above (see
Section~\ref{sec:model_lsf}), the instrumental profile of the FPI can be
well described by the Lorentz function. At the same time, the LSF shape
of classical integral field spectrographs have an LSF is close to Gaussian (e.g.,
MaNGA \citep{Law2021}, MUSE \citep{MUSE}, IFU within SCORPIO-2
\citep{Afanasiev2018}). We considered four types of LSF in our analysis:

\begin{list}{}{
\setlength\leftmargin{4mm} \setlength\topsep{2mm}
\setlength\parsep{0mm} \setlength\itemsep{2mm} }
\item 1)~Gaussian LSF with the profile width $FWHM_{\rm LSF}$. In this
case the effect of instrumental broadening on the integrated line
profile is similar to that due to turbulent line broadening, so we
consider the quantity $\sqrt{\sigma_{\rm ISM}^2+\sigma_{\rm LSF}^2}$,
where $\sigma_{\rm LSF}=FWHM_{\rm LSF}/(2\sqrt{2\ln 2})$, as a parameter
for further analysis. Here we estimate the ``observable'' velocity
dispersion $\sigma_{\rm obs}$ measured from the integrated spectrum of
the shell by fitting the line profile to a Gaussian function.
\item 2)~The case of the LSF in the form of a Lorentz function. In that
case we set $\sigma_{\rm LSF}=0$ in all further computations. Note that
here the measured velocity dispersion $\sigma_{\rm obs}$ inferred by
fitting the observed profile to a Voigt function is free from the
contribution of instrumental broadening, because it refers to the
Gaussian component \citep{Moiseev2008}. We considered three values of
the Lorentzian profile width corresponding to the three FPIs used in
SCORPIO-2 on the 6-m telescope of the SAO RAS:
\begin{list}{$\bullet$}{
\setlength\leftmargin{6mm}
\setlength\topsep{2mm}
\setlength\parsep{0mm} \setlength\itemsep{2mm} }
\item $FWHM_{\rm LOR}=18$~km\,s$^{-1}$ (for IFP751); \item $FWHM_{\rm
LOR}=36$~km\,s$^{-1}$ (for IFP501);
\item $FWHM_{\rm LOR}=78$~km\,s$^{-1}$ (for IFP186).
\end{list}
\end{list}

\begin{figure*} 
\includegraphics[scale=0.55]{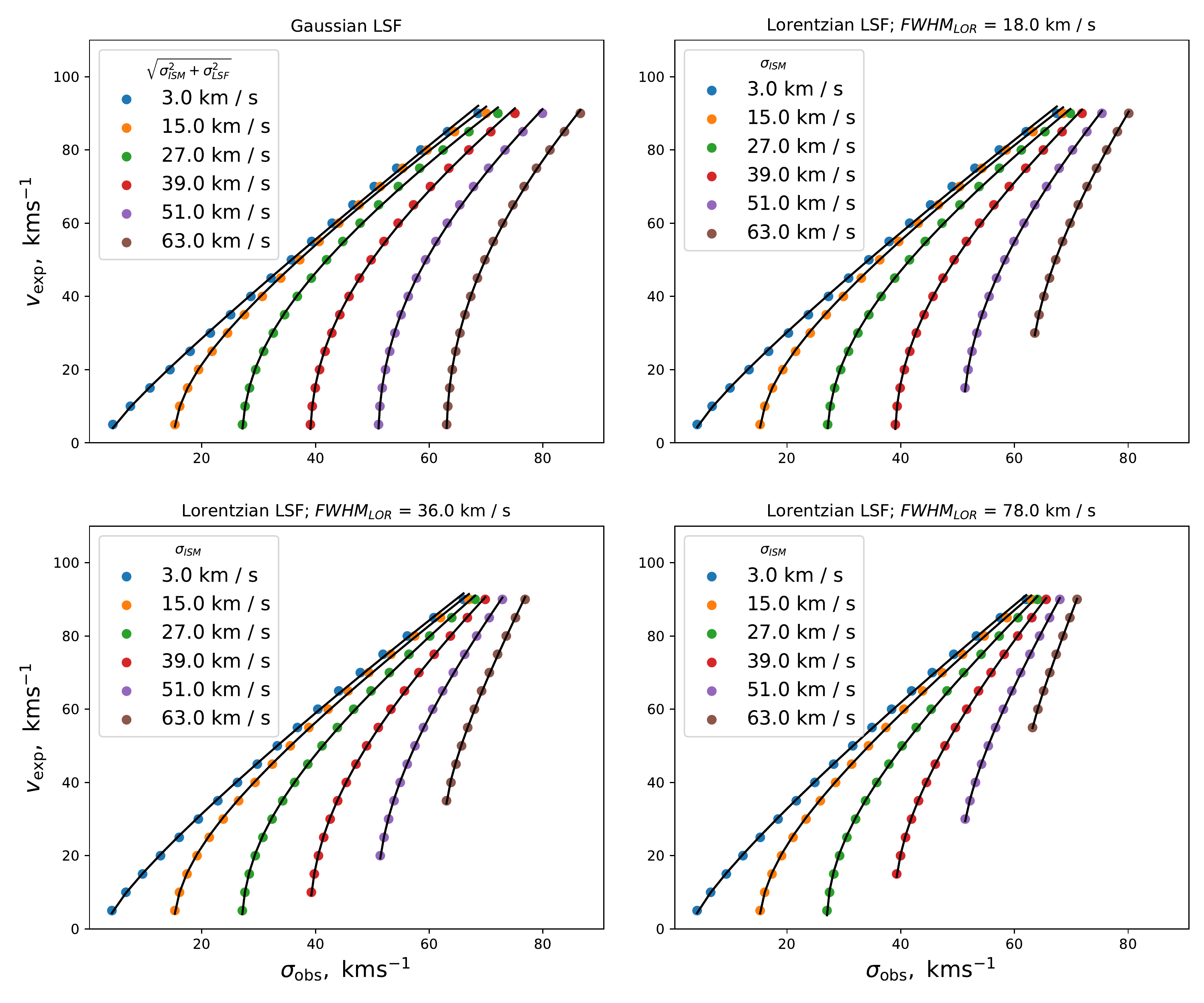}
\caption{Relationship between the expansion velocity $v_{\rm exp}$,
observed velocity dispersion $\sigma_{\rm obs}$, unperturbed
interstellar medium velocity dispersion $\sigma_{\rm ISM}$, and the
shape and width of the instrumental profile ($FWHM_{\rm LOR}$,
$\sigma_{\rm LSF}$) as inferred in terms of the uniform expanding bubble
model. The solid lines show the result of the best approximation by a
relation of the form~(\ref{eq:vexp}).} \label{fig:relation}
\end{figure*}

Fig.~\ref{fig:relation} shows the dependence of the model-estimated
expansion velocity of a bubble on the velocity dispersion $\sigma_{\rm obs}$
measured from its integrated spectrum, for different values
of $\sigma_{\rm ISM}$. For each of these, the resulting dependencies are
well approximated by a relation of the form
\begin{equation}
v_{\rm exp}=k(\sigma_{\rm obs}^2-(\sigma_{\rm ISM}^2+\sigma_{\rm
LSF}^2))^a+v_0. \label{eq:vexp}
\end{equation}

\begin{figure*} 
\includegraphics[scale=0.55]{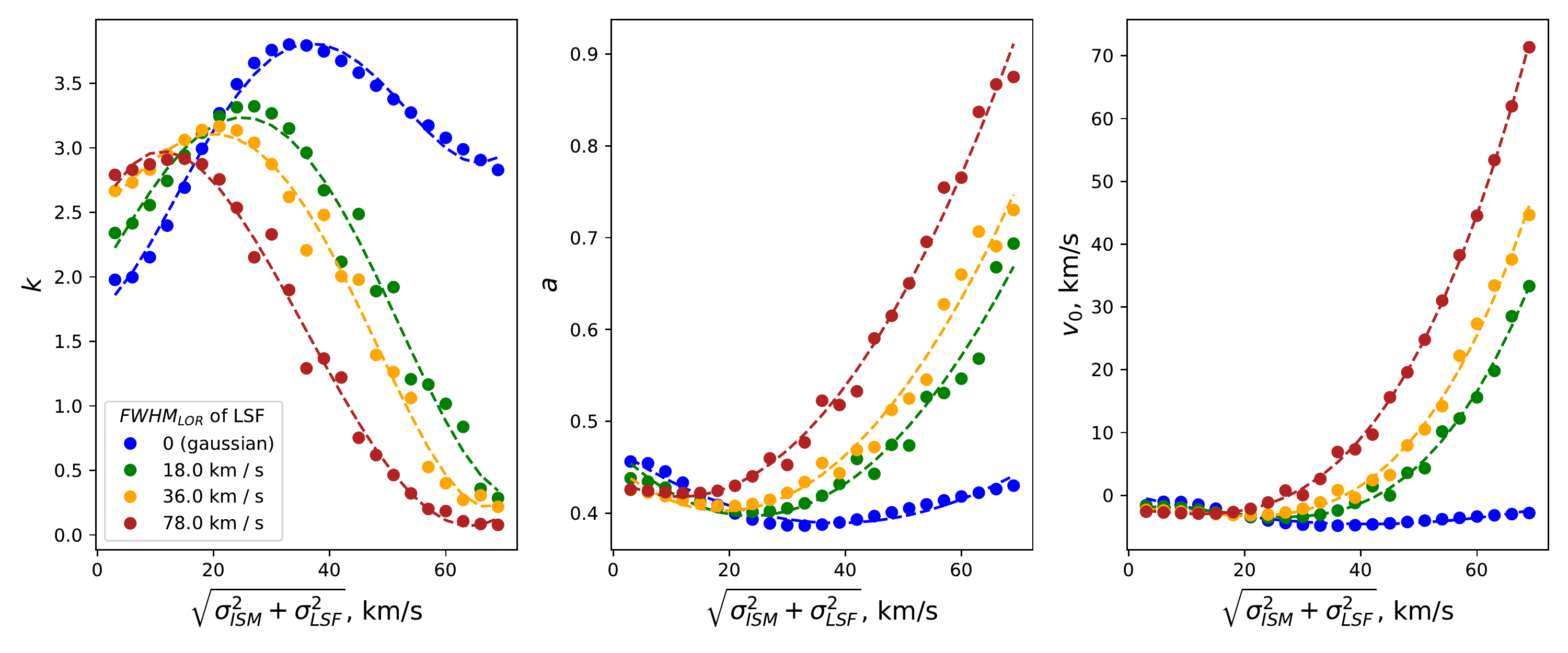}
\caption{Dependence of parameters $k$, $a$ and $c$ in
equation~(\ref{eq:vexp}) on the shape and width of the instrumental
spectrograph contour $\sigma_{\rm LSF}$, $FWHM_{\rm LOR}$ and on the
average gas velocity dispersion $\sigma_{\rm ISM}$ in the galaxy. The
dashed line shows the result of the best polynomial approximation
according to the equations~(\ref{eq:cf_k})--(\ref{eq:cf_v}) and
Table~\ref{tab:coeffs}.} \label{fig:coefficients}
\end{figure*}

The coefficients $k$, $v_0$ and exponent $a$ in the formula depend on
the average dispersion of gas velocities in the galaxy $\sigma_{\rm
ISM}$ as well as on the form and width of the instrumental profile of
the spectrograph (parameters $\sigma_{\rm LSF}$ and  $FWHM_{\rm LOR}$).
Figure~\ref{fig:coefficients} shows the behavior of the dependence of
each parameter on these quantities. Note that the parameter $v_0$ is
relevant only in the case of high velocity dispersion $\sigma_{\rm ISM}$
of unperturbed gas and is most important in the case of the use of the low
spectral resolution FPI, which is poorly suited to study gas
velocity dispersion variations in galaxies. We can assume $v_0\simeq 0$
in most cases of typical velocity dispersions $\sigma_{\rm ISM}\sim
12$--$35$~km\,s$^{-1}$ in nearby galaxies \citep{Moiseev2015}.

The variation of each of the parameters $k$, $a$, $v_0$ in
formula~(\ref{eq:vexp}) as a function of $\sqrt{\sigma_{\rm
ISM}^2+\sigma_{\rm LSF}^2}$ can be described by the following
polynomial relations
\begin{eqnarray} k(x)&=&c_{k4}x^4+c_{k3}x^3+c_{k2}x^2+c_{k1}x+c_{k0},
\label{eq:cf_k}
\\ a(x)&=&c_{a2}x^2+c_{a1}x+c_{a0},\label{eq:cf_a}
\\ v_0(x)&=&c_{v3}x^3+c_{v2}x^2+c_{v1}x+c_{v0}, \label{eq:cf_v}
\end{eqnarray},
where $x=\sqrt{\sigma_{\rm ISM}^2+\sigma_{\rm LSF}^2}$
($\sigma_{\rm LSF}=0$ in the case of the Lorentz instrumental profile). The
coefficients in equations~(\ref{eq:cf_k})--(\ref{eq:cf_v}) are listed in
Table~\ref{tab:coeffs} for all the four LSF types considered.

\begin{table*}
\caption{Polynomial expansion coefficients of parameters $k$, $a$ and
$v_0$ according to equations~(\ref{eq:cf_k})--(\ref{eq:cf_v}) for
different instrumental contour widths $FWHM_{\rm
LOR}$}\label{tab:coeffs}
\centering
\medskip
\begin{tabular}{c|c|c|c|c}
 \hline
 & \multicolumn{4}{c}{$FWHM_{\rm LOR}$} \\
 \cline{2-5}
Parameter & 0~km\,s$^{-1}$ & 18~km\,s$^{-1}$ & 36~km\,s$^{-1}$ & 78~km\,s$^{-1}$ \\
 &  (Gaussian LSF) & (IFP751) & (IFP501) & (IFP186) \\
  \hline
$c_{k0}$ & $1.74$      & $2.0$       & $2.49$      & $2.45$      \\
$c_{k1}$ & $0.0257$    & $0.0741$    & $0.0433$    & $0.102$     \\
$c_{k2}$ & $4.60e-03$  & $4.29e-04$  & $5.99e-04$  & $-5.57e-03$ \\
$c_{k3}$ & $-1.42e-04$ & $-7.34e-05$ & $-7.62e-05$ & $6.37e-05$  \\
$c_{k4}$ & $1.06e-06$  & $6.75e-07$  & $7.48e-07$  & $-1.65e-07$ \\
 \hline
$c_{a0}$ & $0.471$     & $0.472$     & $0.452$     & $0.438$     \\
$c_{a1}$ & $-4.27e-03$ & $-6.30e-03$ & $-5.26e-03$ & $-3.48e-03$ \\
$c_{a2}$ & $5.54e-05$  & $1.33e-04$  & $1.38e-04$  & $1.50e-04$  \\
 \hline
$c_{v0}$ & $0.139$     & $-2.0$      & $-2.17$     & $-2.18$     \\
$c_{v1}$ & $-0.23$     & $0.0922$    & $0.0215$    & $-0.091$    \\
$c_{v2}$ & $2.81e-03$  & $-0.0128$   & $-9.37e-03$ & $-0.0011$   \\
$c_{v3}$ & $-3.18e-07$ & $2.73e-04$  & $2.78e-04$  & $2.61e-04$  \\
\hline
\end{tabular}
\end{table*}

\begin{figure*}
\includegraphics[scale=0.55]{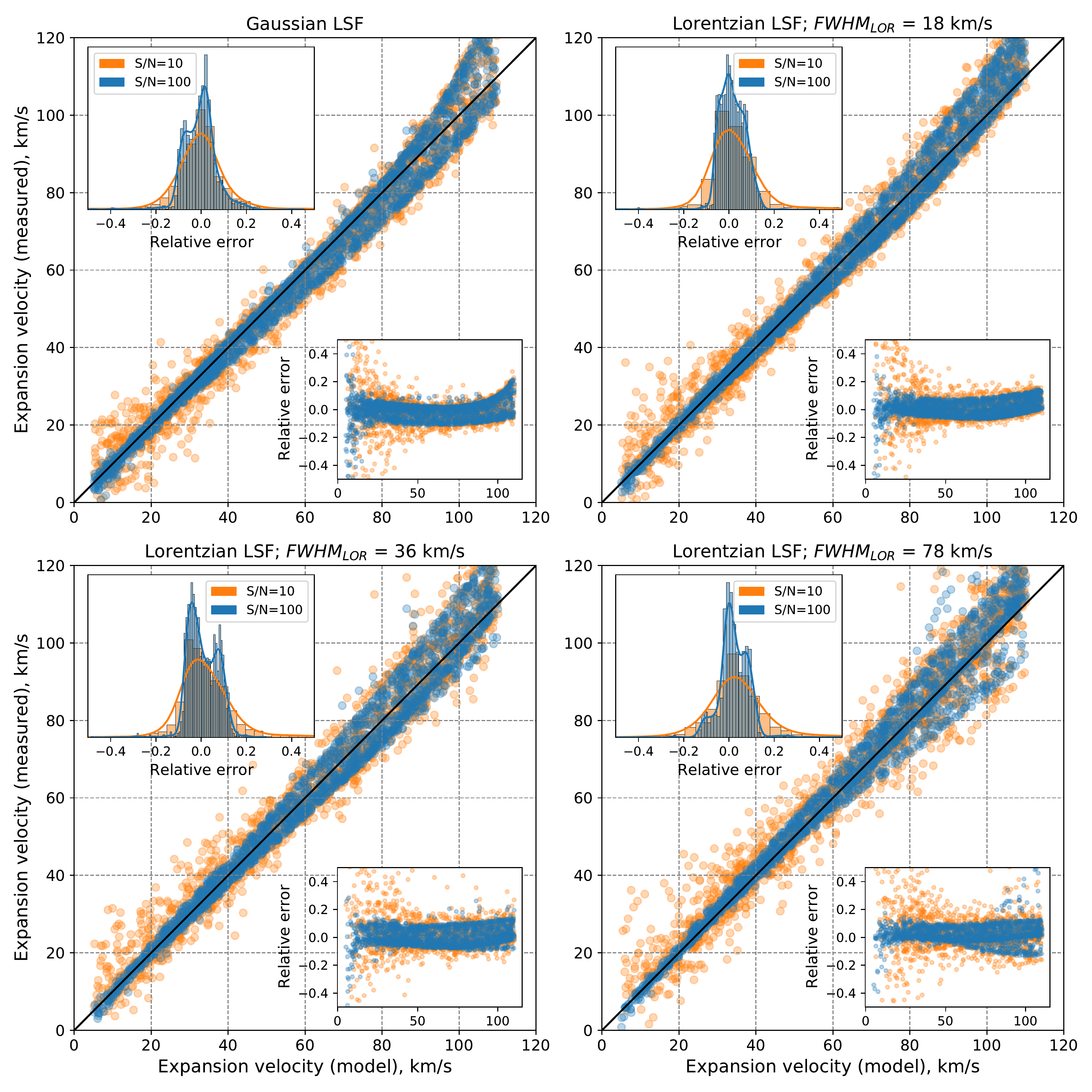}
\caption{Dependence of uniform shell expansion velocity measured by the
equations~(\ref{eq:vexp})--(\ref{eq:cf_v}) on the input velocity given in
the simulation. For each LSF type and width, model line profiles are
generated for 4000 bubbles with
$S/N=10$ (the orange symbols) and
$S/N=100$ (the blue symbols),
expansion velocity $v_{\rm exp}$
(5 to 110~km\,s$^{-1}$), ambient gas velocity dispersion
$\sigma_{\rm ISM}$
(10 to 70~km\,s$^{-1}$,
it includes the instrumental profile width in the
case of Gaussian LSF). Panels for different LSF types and widths are
presented. The histogram in each panel shows the distribution of the
relative measurement error
\mbox{$(v_{\rm exp}[\mathrm{measured}]-v_{\rm
exp}[\mathrm{model}])/v_{\rm exp}[\mathrm{model}]$}.
The bottom of each panel shows the distribution
of the relative error on input $v_{\rm exp}$.}
\label{fig:mc_test}
\end{figure*}

To estimate the relative error of the shell expansion velocity inferred
from equation~(\ref{eq:vexp}) and the robustness of the
equations~(\ref{eq:vexp})--(\ref{eq:cf_v}) with respect to noise, we
performed an additional Monte Carlo simulation. For each of the four LSF
types, we simulated 4000 spectra of bubbles with random parameters
constrained in the intervals \mbox {$\sqrt{\sigma_{\rm
ISM}^2+\sigma_{\rm LSF}^2}=10$--$70$}~km\,s$^{-1}$ and \mbox {$v_{\rm
exp}=5$--$110$}~km\,s$^{-1}$, and added noise with final signal-to-noise
ratios of $S/N=10$ and \mbox {$S/N=100$}. We applied equations~(\ref{eq:vexp})--(\ref{eq:cf_v})
to the resulting synthetic profiles
and estimated the bubble expansion velocity in each case. A comparison of
the resulting estimate with its true value set in the model demonstrates
the accuracy of the technique. Fig.~\ref{fig:mc_test} shows the
dependence of these two quantities on each other as well as the
distribution of the relative measurement error computed as $(v_{\rm
exp}[\mathrm{measured}]-v_{\rm exp}[\mathrm{model}])/v_{\rm
exp}[\mathrm{model}]$. As is evident from the figure, the technique
described above is quite robust with respect to noise and the shell
expansion velocity to be estimated with an accuracy of about 10\% over a
wide range of velocities.

The method described above is based on the uniform bubble model, but real
H\,II regions have a more complex morphology. First, the observed
superbubbles in extragalactic star-forming regions form as a result of
energy inflow from multiple massive stars, i.e., in fact, they are complexes
of merging local bubbles at different stages of evolution. As was
shown by \citet{TT1996}, the integrated spectrum of such unresolved
complexes ca still be well described by the model of a single uniform superbubble,
but there may be an additional broad low-intensity component in the line
wings---its contribution does not affect the analysis considered in
this paper. The second important point is that real H\,II regions are
not uniform but highly clumped with significant local density and
temperature inhomogeneities. Let us check to what extent this effect
affects our estimates.

\begin{figure}
\includegraphics[scale=0.55]{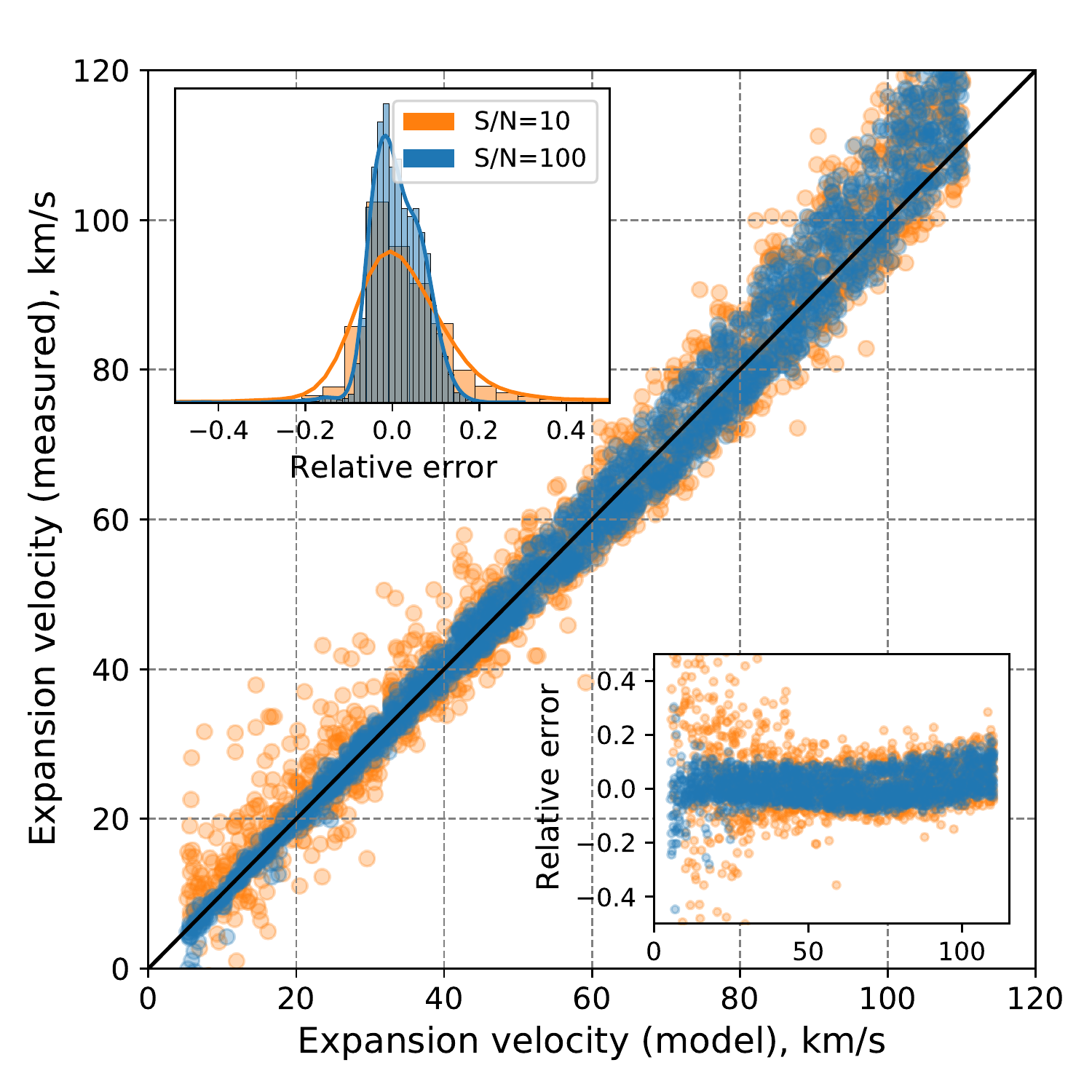}
\caption{Same as in Fig.~\ref{fig:mc_test}, but for the models of bubbles with
nonuniform brightness distribution. The dependences are shown only for
the LSF in the form of a Lorentz function with profile width $FWHM_{\rm
LOR}=18$~km\,s$^{-1}$.} \label{fig:nonisohist}
\end{figure}

We now consider a shell in the form of a thin expanding sphere, but this
time the brightness of each element $dV$ in Fig.~\ref{fig:isosphere}
varies with angles $\theta$ and $\varphi$. For this purpose, we set the
distribution $I(\theta, \varphi)$ in the form of the sum of spherical
harmonics up to and including the eighth order with random amplitudes
and with the dominant contribution of the angle-independent (zero)
component. The final intensity distribution from each element in the
bubble is a set of separate spots of various sizes with higher- and
lower-than average brightness. As in the case of the uniform model, the
integrated spectrum of the resulting shell is convolved with a Gaussian to
account for thermal and turbulent broadening and with a Gaussian or
Lorentz function describing the instrumental profile of the
spectrograph. To test the applicability of equations~\mbox
{(\ref{eq:vexp})--(\ref{eq:cf_v})} to such structures with non-uniform
brightness distributions, we performed Monte Carlo simulations similar
to those described above, but this time with synthetic profile generated
in terms of a nonuniform bubble model. Fig.~\ref{fig:nonisohist} shows
the result of comparing the measured and modeled bubble expansion
velocities (for $FWHM_{\rm LOR}=18$~km\,s$^{-1}$). The number of regions
exhibiting a significant deviation of the result from the model is
slightly higher than in the case of a uniform shell, but on the average
the relative error is still at the level of about 10\%. Hence the method
described is also applicable in the case of shell structures with
nonuniform brightness distribution.

\begin{figure} 
\includegraphics[scale=0.55]{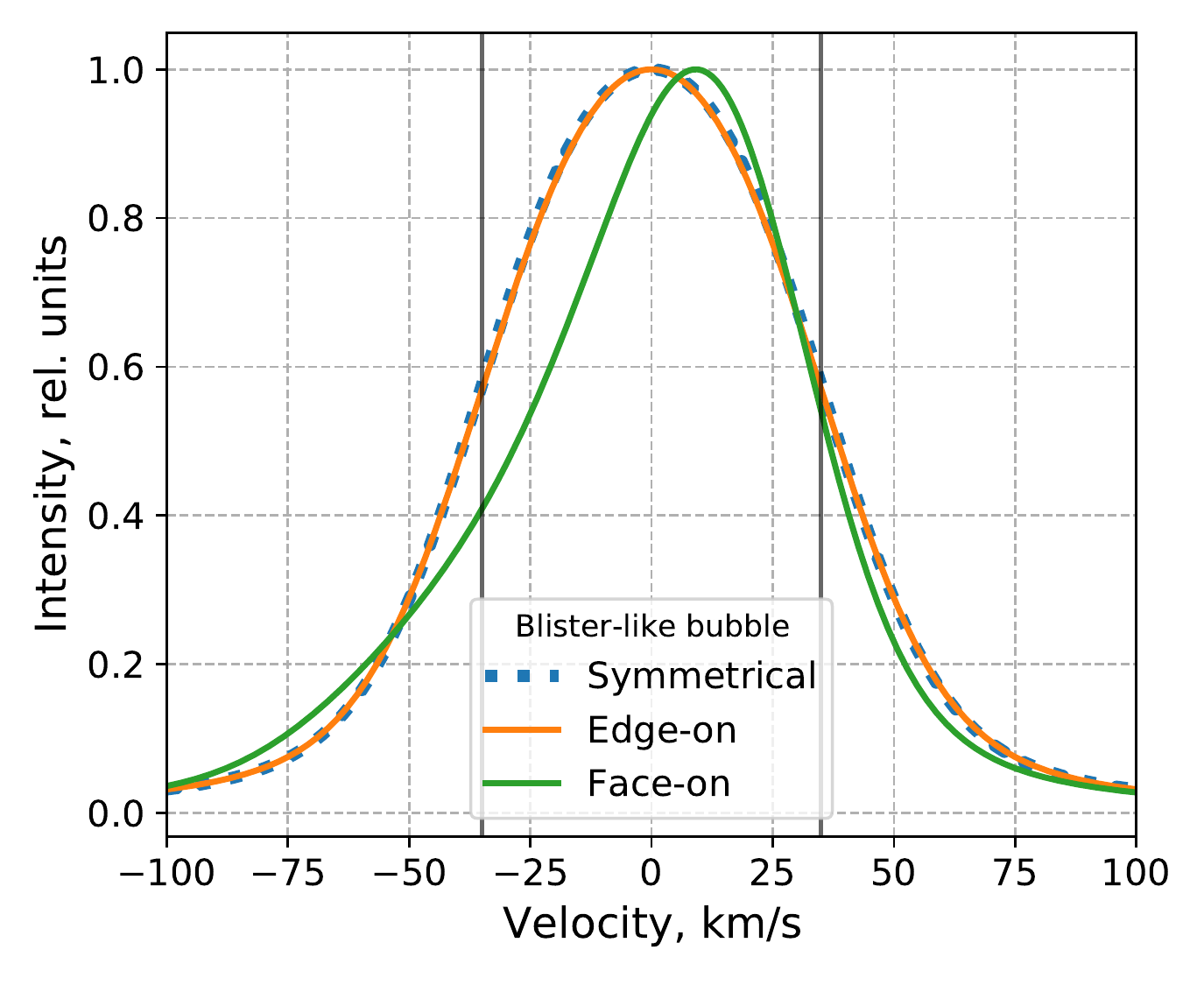}
\caption{Integrated model line profile from an expanding spherical bubble
(the blue line) in the form of a hemisphere at different orientations.
In the case where the bright part of the bubble is in the sky plane (the
green line), the resulting integrated profile is significantly
asymmetric. When the bright part of the bubble is perpendicular to the
sky plane (the orange line) the resulting line profile is
indistinguishable from the case of spherical symmetry. The vertical
lines correspond to the preset expansion velocity \mbox {$v_{\rm
exp}=35$}~km\,s$^{-1}$.}  \label{fig:blister}
\end{figure}

In the cases where star formation occurs at the boundary of a molecular
cloud, significant gas density gradient of gas density around massive
stars can cause the formation of asymmetric shells with azimuthal
gradient of brightness and velocity. Typical examples are blister-type
nebulae (see, e.g., \citet{Egorov2010}). To test the applicability of
the above technique to such objects, we simulated a shell structure with
smooth azimuthal variation of brightness and velocity. Given that the
brightness of the emission-line nebula depends on gas density $I\sim
n_e^2 \sim \rho^2$ \citep{Osterbrock2006}, and that relation $v \rho^2
\sim const$ \citep{McKee1975} is usually fulfilled for radiative shocks,
we can specify the expansion velocity of our asymmetric shell at each
point by the law $v \sim I^{-1}$. From observational point of view we can
distinguish two extreme cases: the bright (dense) side of the shell is
in the sky plane (face-on) or perpendicular to it (edge-on). In the
first case such a shell appears as an ordinary symmetric H\,II region,
whereas in the second case there should be clear brightness gradient (in
the case of sufficient angular resolution). Fig.~\ref{fig:blister} shows
the simulated integrated profiles for each of these cases. As is evident
from the figure, the integrated line profile of a blister-type profile is
indistinguishable from the spherically symmetric case in its
edge-on orientation, and hence the method we have described above is
fully applicable to such objects. On the other hand, in the case of
face-on orientation the integrated profile is significantly asymmetric,
resulting in underestimated measured velocity dispersion and,
consequently, underestimated bubble expansion velocity. Thus, in the case
shown in Fig.~\ref{fig:blister}, a formal fit of the asymmetric face-on
profile of the shell by a single-component Voigt function and
application of equations~(\ref{eq:vexp})--(\ref{eq:cf_v}) yields an
expansion velocity estimate of \mbox {$v_{\rm exp}=30$}~km\,s$^{-1}$
instead of 35~km\,s$^{-1}$ value incorporated into the model, and the
discrepancy is significantly greater in the case of higher expansion
velocities. Unfortunately, even in the case of sufficiently high angular
resolution distinguishing such face-on blister-type nebulae from
spherically symmetric ones is by no means a trivial task. In this
connection, it is worth noting that the method presented in this paper
may be unreliable in cases where the integrated line profile of a nebula
is significantly asymmetric.

The method described in this section allows us to use
equations~(\ref{eq:vexp})--(\ref{eq:cf_v}) to determine the expansion
velocities of spatially unresolved (or poorly resolved) bubbles in the
interstellar medium of galaxies based on the gas velocity dispersion
$\sigma_{\rm obs}$ measured from the integrated spectrum of a bubble and the
average gas velocity dispersion in the galaxy $\sigma_{\rm ISM}$. The
gas velocity dispersion $\sigma_{\rm ISM}$ in this case can be inferred,
for example, from the estimated flux-weighted average gas velocity
dispersion of the bright H\,II regions in the galaxy, which stand out in
the \mbox {($I$--$\sigma$)} diagrams. The presented method can also be
useful in the case of spatially resolved shell structures when
insufficient spectral resolution prevents the decomposition
of the profile into kinematically isolated components. In the next
section we test the method on real observational data.

\section{THE  GALAXY IC\,1613: TESTING THE METHOD ON OBSERVATIONAL
DATA}\label{sec:obs}

\begin{figure*} 
\includegraphics[scale=0.55]{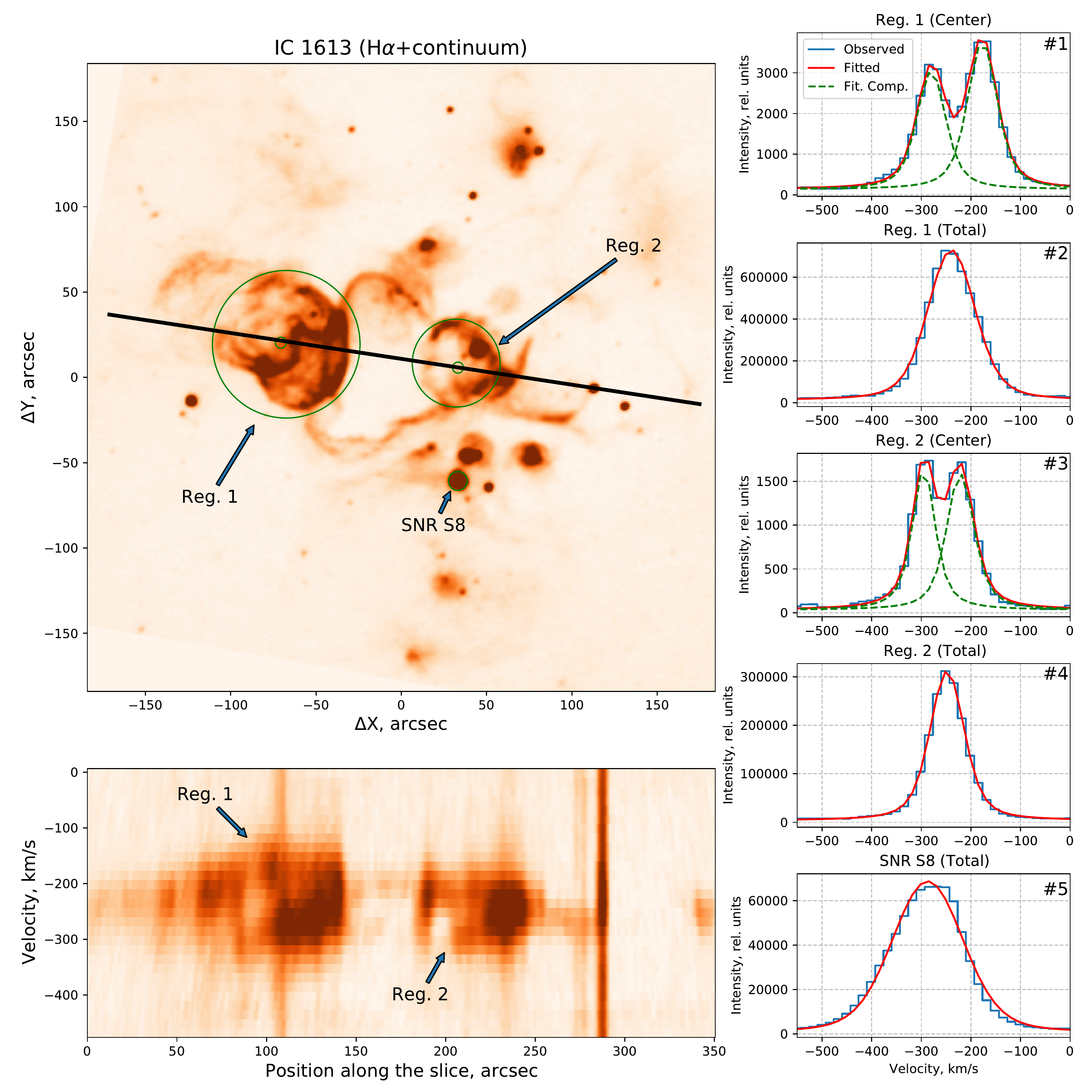}
\caption{Complex of superbubbles in the galaxy IC\,1613 according to
H$\alpha$-line SCORPIO observations on the 6-m telescope of the SAO RAS. Top left
panel: H$\alpha$-line image of the region, the position of the cut (the black line)
for the ``position-velocity'' diagram (shown in the bottom
left panel). The right panel shows the observed H$\alpha$-line profiles
by integration in the apertures shown by the green circles in the image.
The profiles \#1, \#3 are obtained for the central part of the
considered regions, and profiles \#2, \#4---for the entire region.
Profile \#5 corresponds to the bubble surrounding the young  SNR\,S8
supernova remnant. The dashed green line shows the individual
components, and the red line, the final fit of the observed profile by a
one- or two-component Voigt function.}     \label{fig:ic1613}
\end{figure*}

Nearby star-forming dwarf galaxies are an excellent laboratory for
studying the interaction processes of massive stars and the interstellar
medium. Because of the lack of spiral density waves and the thick
gaseous disk extended supershell complexes are often observed in the
interstellar medium of such galaxies. Such a pattern can be clearly seen
in IC\,1613---a nearby Local Group galaxy (at a distance of
$D\sim760$~kpc \citep{Karach2013}. A complex of expanding superbubbles of
ionized gas extending over about 1~kpc is observed in its eastern part
(see Fig.~\ref{fig:ic1613}). This complex of superbubbles has been studied in
detail previously by \citet{Lozinskaya2003} and is well suited for
testing the method described in Section~\ref{sec:method_apply} on real
data.

We use H$\alpha$-line data obtained on September 13, 2001 with the FPI
(IFP501) of the SCORPIO focal reducer \citep{scorpio} mounted on the BTA
6-m telescope of the SAO RAS. For a detailed description of the observations and
data processing see \citet{Lozinskaya2003}.

As is evident from Fig.~\ref{fig:ic1613}, the two regions labeled Reg.~1
and Reg.~2 exhibit a clear shell-like morphology and reveal a characteristic
velocity ellipse in the ``position-velocity'' diagrams, indicating the
observed superbubble expansion. The angular size of these superbubbles is on the
order of $80\arcsec$ and $55\arcsec$ (290 and 200~pc), respectively,
which significantly exceeds the angular resolution of the observational
data (about $1.5\arcsec$). The observed H$\alpha$ profile in the central
part of the shells (\#1 and \#3 in Fig.~\ref{fig:ic1613}) is clearly
separated into components corresponding to the approaching and the
receding sides. The expansion velocities of the Reg.~1 and Reg.~2 shells
computed from the maximum separation of the H$\alpha$ profile components
are equal to $v_{\rm exp}=51$~km\,s$^{-1}$ and 37~km\,s$^{-1}$
respectively.

Superbubbles similar to Reg.~1 and Reg.~2 observed in more distant galaxies
will have smaller angular sizes, and they should be practically
unresolvable at $D\sim10$~Mpc. In such a case their integrated spectrum
will be observed, as in the case of profiles \#2 and \#4 in
Fig.~\ref{fig:ic1613}. No separation into individual components can be
observed, but the H$\alpha$ profile is broadened. The measured velocity
dispersion (free of instrumental broadening) is \mbox{$\sigma_{\rm
obs}=39$}~km\,s$^{-1}$ and 26~km\,s$^{-1}$ for Reg.~1 and Reg.~2,
respectively. We can estimate the velocity dispersion of the surrounding
interstellar medium as the flux-weighted average H$\alpha$-line
velocity dispersion in regions with intensities higher than
the median I\,(H$\alpha$) in the galaxy (to exclude the contribution of
diffuse ionized gas). The inferred velocity dispersion \mbox{$\sigma_{\rm ISM}=15$}~km\,s$^{-1}$
is consistent with the estimate of
the velocity dispersion of individual components on the profile \#3, as well as
with an estimate for separate bright H\,II regions, which exhibit no
signs of expansion. We can adopt these values $\sigma_{\rm ISM}$ and
$\sigma_{\rm obs}$ and $FWHM_{\rm LOR}=36$~~km\,s$^{-1}$ for IFP501 and
use equations (\ref{eq:vexp})--(\ref{eq:cf_v}) to estimate expansion
velocity by the method proposed in this paper. The inferred expansion
velocity estimates \mbox {$v_{\rm exp}=53$}~km\,s$^{-1}$ and
33~km\,s$^{-1}$ for Reg.~1 and Reg.~2, respectively, agree well with
those found above, which are based on the H$\alpha$-line profile
decomposition within the previously quoted relative error of the method
(on the order of 10\%).

The only known supernova remnant in the IC\,1613 galaxy---SNR\,S8---is also
observed toward the complex considered, but, unlike Reg.~1
and Reg.~2 regions, it is practically unresolved, making it impossible
to directly measure directly its expansion velocity. Observational data
obtained with MPFS spectrograph attached at the 6-m telescope of
the SAO RAS
\citet{Lozinskaya1998} revealed no velocity dispersion gradient that
could clearly indicate SNR expansion, and yielded a rough estimate
$v_{\rm exp}\sim 50$--$100$~km\,s$^{-1}$ based on the variation of the
observed radial velocity toward the SNR ($v_{\rm exp}<300$~km\,s$^{-1}$
if weak emission structures are also considered). A detailed study of
SNR\,S8 is also given in \citet{Schlegel2019}. Direct measurements of
the expansion velocity of the supernova remnant are still lacking, but
new high-resolution X-ray data have allowed the authors to estimate
SNR\,S8 properties, including its age, independent of kinematic
parameters.

We now try to estimate the expansion velocity of SNR\,S8 and its
kinematic age from available observational data obtained with the FPI
using equations~(\ref{eq:vexp})--(\ref{eq:cf_v}). The integrated profile
in the H$\alpha$ line (\#5 in Fig.~\ref{fig:ic1613}) is significantly
broadened and slightly asymmetric. The measured velocity dispersion
\mbox {$\sigma_{\rm obs}=64$}~km\,s$^{-1}$ corresponds to the optical
shell expansion velocity $v_{\rm exp}=84$~km\,s$^{-1}$, which agrees
with the approximate estimate obtained in \citet{Lozinskaya1998}. The
velocity thus measured corresponds to the velocity of the radiative
shock propagating in regions of increased supernova remnant density,
while the shock velocity outside dense clouds, which determines the
remnant expansion velocity, is related to the gas density as
\citep{McKee1975}: \mbox{$\beta n_i v_s^2 = n_c v_c^2$}, where
$\beta=3.15$ for fast shocks, $n_i$ and $n_c$ are equal to the gas
density in unperturbed interstellar medium and in a dense cloud,
respectively; $v_s$ is the SNR shock front velocity in unperturbed medium,
$v_c$---radiative shock velocity in a dense cloud. When applied to SNR\,S8
$v_c = v_{\rm exp} \simeq 84$~km\,s$^{-1}$, $n_i \simeq 1.6$~cm$^{-3}$
(according to \citet{Schlegel2019}), \mbox {$n_c=n_e/1.2 \simeq 1080$}
(according to electron density estimate $n_e$ obtained by
\citet{Lozinskaya1998} by the [S\,II]\,6717/6731~\AA\ lines ratio). From this,
we find the velocity of the shock front SNR\,S8 $v_s \simeq
1230$~km\,s$^{-1}$. Assuming that SNR\,S8 is at the adiabatic stage of
expansion (similar to \citet{Schlegel2019}), we can describe its
evolution as $R_s \sim t^{0.4}$ \citep{Sedov}, which means that
$t=0.4R_s/v_s$. Taking the estimate $R_s\sim9.5$~pc obtained in
\citet{Schlegel2019} from X-ray images, we determine the kinematic age
value of SNR\,S8 \mbox {$t\simeq3100$}~years, which agrees well with the
independent estimate \mbox {$t\simeq3380$--$5650$}~years obtained in
\citet{Schlegel2019} who did not use an information on the gas kinematics.

We thus used the IC\,1613 galaxy as an example to demonstrate that the
method for measuring a bubble/superbubble expansion velocity described in this
paper is applicable to real observational data and yields the expansion
velocity and age estimates that are consistent
with other independent measurements.

\section{SUMMARY}\label{sec:summary}

Integral field spectroscopy reveals variations of the observed gas velocity
dispersion in galaxies that are driven by presence  of the expanding bubbles and superbubbles
surrounding massive stars and clusters in star-forming regions
\citep[see, e.~g.,][]{Moiseev2012}. With high enough spectral resolution
the observed emission line profile in such regions can be
decomposed into kinematically isolated components allowing the bubble
expansion velocity to be measured (see, e.g., \citet{Lozinskaya2003,
Egorov2014, Egorov2017, Oparin2020}), but this is rarely possible for
galaxies located beyond the Local Group because of limited spatial
resolution. In this paper, we propose a method allowing one to overcome
this limitation and estimate the shell expansion velocity from the
dispersion of gas velocities measured from integrated spectrum.

We considered an analytical model of a uniform thin shell expanding in a
turbulized interstellar medium and found how thermal, turbulent, and
instrumental broadening and bubble expansion affect the shape of the
observed emission line profile. We used this model to demonstrate that
the shell expansion velocity obtained by decomposing the profile into
kinematically separated components can be significantly underestimated
if the angular resolution is insufficient. This results in overestimated
kinematic age and underestimated energy inflow required for the
formation of the shell, and these biases can affect the estimated
efficiency of the transfer of energy from stars to kinetic energy of
superbubbles when comparing observations and theoretical models.

The models we have built showed that the bubble expansion velocity can be
determined by equation~(\ref{eq:vexp}), which includes the gas velocity dispersion
of the bubble (measured from the integrated spectrum) and
in the surrounding unperturbed interstellar medium (which can be
estimated for each galaxy as a flux-weighted average value for bright
H\,II regions). We showed that the parameters used in
equation~(\ref{eq:vexp}) depend on the average gas velocity dispersion
in the galaxy and can be approximated by
polynomials~(\ref{eq:cf_k})--(\ref{eq:cf_v}). The coefficients in these
polynomials depend on the type of instrumental profile of the
spectrograph. We considered the cases of LSF in the form of Gaussian
functions (applicable to most classical integral field spectrographs) with
arbitrary spectral resolution and Lorentz functions (for FPI) with
the resolution corresponding to IFP751, IFP501, and IFP186, which are
used in SCORPIO-2 instrument at the 6-m telescope of the
SAO RAS.
Table~\ref{tab:coeffs} lists the coefficients inferred for these cases.

We applied the method described to synthetic spectra of bubbles having
random parameters and different $S/N$ ratios in their spectra and showed
that the relative error of the measurement of the bubble expansion
velocity is on average about 10\%. To further test the applicability of
the method, we also analyzed a bubble model with a non-uniform brightness
distribution and found the result to be the same. We showed that the
method is also applicable to bubbles with large-scale inhomogeneities
(for example, those having the form of a hemisphere), but in this case,
the spatial orientation of bright regions plays an important role.

We used the method described to estimate the expansion velocities of
well-resolved shells of ionized gas and of the only known supernova
remnant (SNR\,S8) in the nearby galaxy IC\,1613. The estimates obtained
for the superbubbles agree well with measurements made by decomposing the
line profile into kinematically separated components. For the first time we estimated
the kinematic age of SNR\,S8 (\mbox {$t\simeq 3100$}~years)
and our result agrees well with the age obtained earlier from X-ray
data \citep{Schlegel2019}.

The method proposed in this paper allows measuring the expansion
velocities of spatially unresolved (or poorly resolved) superbubbles in
galaxies outside the Local Group. In particular, it is intended to be
used to measure the expansion velocities, kinematic ages, and energy of
superbubbles identified by increased gas velocity dispersion from
observations of nearby dwarf galaxies with a FPI at the 6-m
telescope of the SAO RAS within the framework of SIGMA-FPI
catalog\footnote{\url{http://sigma.sai.msu.ru}}. (Egorov et al.,
in prep.).

\begin{acknowledgements}
The authors are grateful to A.~V.~Moiseev for
the discussion on the preliminary results of the work and to the anonymous
referee for valuable comments.
\end{acknowledgements}

\section*{FUNDING}
This work was supported by the Russian Science
Foundation (Grant No.~19-72-00149). Observations on the 6-m telescope of
the Special Astrophysical Observatory of the Russian Academy of Sciences
are supported by the Ministry of Science and Higher Education of the
Russian Federation (contract No. 05.619.21.0016, project identifier
RFMEFI61919X0016).  The renovation of telescope equipment is
currently provided within the national project ''Science.''

\section*{CONFLICT OF INTEREST}
The authors declare that there is no
conflict of interest.


\begin{flushright}
{\it Translated by A.~Dambis}
\end{flushright}

\end{document}